\documentclass[12pt]{article}
\usepackage{epsfig}
\usepackage{amsfonts} % para incluir fonts y
\usepackage{amssymb}  % simbolos de la AMS
\usepackage{graphicx} % standard LaTeX graphics tool
\usepackage{amsmath}  % for including eps-figure files
\usepackage{amsmath}  % paquete matemÃÂ¡tico
\usepackage[latin1]{inputenc}
\usepackage{color}
\def\Title#1{\begin{center} {\Large {\bf #1} } \end{center}}

\begin{document}

\Title{BCS-BEC Crossover and the EoS of Strongly Interacting Systems}

\bigskip\bigskip

%+\addcontentsline{toc}{chapter}{{\it L. Skywalker}}
%+\label{SkywalkerLukeStart}

\begin{raggedright}
{\it E. J. Ferrer\index{Ferrer, E}\\
Department of Physics, University of Texas at El Paso,
El Paso, TX 79968, USA\\
 Email: ejferrer@utep.edu\\
}
\bigskip\bigskip
\end{raggedright}

\section{Introduction}

The two extremes of the QCD phase diagram in the temperature-density ($T-\mu_B$) plane are well understood (see \cite{Fukushima} for a review), having the quark-gluon plasma in the high-temperature/low-density corner and the color-superconductivity CFL phase on the opposite one. In those extremes, due to the large energy scales, the coupling constant runs to smaller values giving rise to asymptotic freedom \cite{Politzer}. Hence, it is natural to expect that in the intermediate temperature and density regions a phase transition occurs from a confined to a deconfined phase where gluons and quarks liberate from hadrons.

At present it has been well established that the stable color-superconducting phase at asymptotically large densities is the CFL phase \cite{CFL}. In this case, not only all the quarks are basically massless, but also the electrical and color neutrality of the system are automatically satisfied. However, the next stable phase down in density remains unknown to this date. The problem is related to the appearance of chromomagnetic instabilities at intermediate densities that render the known phases unphysical in this regime.
The origin of the instabilities can be traced back to the separation of the Fermi surfaces of the quarks participating in the pairing at intermediate densities. The pairing stress occurs once the strange quark mass cannot be neglected and the electric and color neutralities constraints are imposed \cite{gCFL}.

Finding the stable superconducting ground state at moderate densities is one of the main questions in the field at present. This is in particular relevant for nuclear astrophysics, as the core of neutron stars will have realistic, intermediate densities, probably large enough for the quark and gluon degrees of freedom to be manifested, but insufficient to be in the stable CFL phase.

One possible scenario where chromomagnetic instabilities can be avoided occurs if in the region of moderate-low densities the strong coupling constant becomes sufficiently high ($G_D\approx G_S\approx 1/\Lambda^2$, with $G_S$ and $G_S$ denoting the diquark and quark-antiquark coupling constants respectively) \cite{Strong-Coupling}. On the other hand, the increase of the coupling constant strength at low density can modify the properties of the ground state as indicated by the significant decrease of the Cooper-pair coherence length, which can reach values of the order of the inter-quark spacing \cite{Coher-length}. As already found in other physical contexts (see \cite{BCS-BEC} for a review), this fact strongly suggests the possibility of a crossover from a color-superconducting BCS dynamics to a BEC one, where although the symmetry breaking order parameter (the diquark condensate) is the same, the quasiparticle spectra in the two regions are completely different. As we showed in \cite{Jason}, in the BCS region, where the diquark coupling is relatively weak, the energy spectrum of the excitations has a fermionic nature, while in the strong-coupling region, formed by the BEC molecules, the energy spectrum of the quasiparticles is bosonic.

In must of the studies of the BCS-BEC crossover in quark matter one important ingredient was left out
up to recently \cite{Wang}: the external magnetic field. However, magnetic
fields are endemic in neutron stars. Pulsars's magnetic fields range
between $10^{12}$ to $10^{13}$ G \cite{Pulsars}, and for magnetars
they can be as large as $10^{14}-10^{15}$ G \cite{Magnetars} on
the surface and presumably much larger in the core. Upper limit estimates
for neutron star magnetic fields indicate that their magnitude can
reach $\sim10^{18}-10^{20}$ G \cite{virial}-\cite{EoS-B2010}. Heavy
ion collisions can also generate very strong magnetic fields produced
in peripheral collisions by the positively charged ions moving at almost the speed of light. As argued in \cite{Kharzaeev}, these strong magnetic fields, produced during the first instants after a collision, can create the conditions for observable QCD effects. These effects can be prominent because the reached magnetic fields are of the order of, or higher than, the QCD scale. There are both theoretical and experimental indications that the colliding charged ions can indeed generate magnetic fields estimated to be of order $eB\sim 2m^2_p$ ($\sim 10^{18}$ G) for the top collision of the order of 200 GeV, in non-central Au-Au collisions at RHIC, or even larger, $eB\sim 15 m^2_p$ ($\sim 10^{19}$ G), at future LHC experiments  \cite{Shokov, Abelev}. Even though these magnetic fields decay quickly, they only decay to a tenth of the original value for a time scale of order of the inverse of the saturation scale at RHIC \cite{Larry}, hence they may influence the properties of the QCD phases probed by the experiment. Strong magnetic fields will likely be also generated in the future planned experiments at FAIR, NICA and JPARK, which will make possible to explore the region of higher densities under a magnetic field.

In this talk I will discuss the implications for the equation of state (EoS) of strongly coupled quark matter of the BCS-BEC crossover, as well as the effect of an applied strong magnetic field on that crossover. The details of the findings I am discussing here can be found in Refs. \cite{Jason, Wang}.

\section{Threshold Coupling for BCS-BEC Crossover}\label{section2}

Here, I want to discuss how to establish a clear criterium to fix the critical value of the diquark interaction strength for the BCS-BEC crossover. For our analysis, we consider a simplified pure fermion system with a four-fermion interaction Lagrangian density \cite{R-BCS-BEC-1},
\begin{equation}\label{lagrangian}
\mathcal{L}=\bar{\psi}(i\gamma^{\mu}\partial_{\mu}+\gamma_0\mu-m)\psi+\frac{g}{4}(\bar{\psi}i\gamma_{5}C\bar{\psi}^T)(\psi^TCi\gamma_5\psi),
\end{equation}
where $C=i\gamma_0\gamma_2$ is the charge conjugation matrix, $m$ the fermion mass, $\mu$ the chemical potential defining the Fermi energy, and $g$ the attractive coupling constant in the $J^P=0^+$ channel that parameterizes the strength of the interaction.

After the Hubbard-Stratonovich transformation we have that the system free energy at finite temperature and in the mean-field approximation with gap parameter $\Delta=\langle g\psi^TCi\gamma_5\psi/2\rangle$, is given by
\begin{equation}\label{MF-potential}
\Omega_T=-\frac{1}{\beta}\sum_{n=0}^\infty\int\frac{d^3k}{(2\pi)^4}~Tr \ln~[\beta G^{-1}(i\omega_n,\textbf{k})]+\frac{\Delta^{2}}{g},
\end{equation}
where $G^{-1}(i\omega_n,\textbf{k})$ is the inverse propagator in Nambu-Gor'kov space in the field basis $\Psi^T=(\psi, \psi_C)$, with $\psi_C=C\overline{\psi}^T$ being the charge-conjugate spinors,
\begin{equation}\label{Propagator}
G^{-1}(i\omega_n,\textbf{k})=(\omega_n +\mu \sigma_3)\gamma_0-\gamma \cdot \textbf{k}-m+i\gamma_5\Delta \sigma_+ +i\gamma_5 \Delta^*\sigma_-
\end{equation}
Here, $\omega_n=(2n+1)\pi/\beta$ are the fermion Matsubara frequencies, and $\sigma_{\pm}=\sigma_1\pm i\sigma_2$, with $\sigma_{1,2}$ denoting the corresponding Pauli matrices. After taking the trace and the sum in Matsubara frequencies in (\ref{MF-potential}) it is obtained in the zero-temperature limit
\begin{equation}\label{T0-potential}
\Omega_{0}=-\sum\limits_{e=\pm 1}\int_{\Lambda}\frac{d^{3}k}{(2\pi)^{3}}\;\epsilon_{k}^{e}+\frac{\Delta^{2}}{g},
\end{equation}
where $\Lambda$ is an appropriate momentum cutoff to regularize the momentum integral in the ultraviolet, and the quasiparticle energy spectrum, $\epsilon_{k}^{e}$, which corresponds to particle ($e=+$) and
 antiparticle ($e=-$) is given by
\begin{equation}\label{spectrum}
\epsilon_{k}^{e}=\sqrt{(\epsilon_{k}-e\mu)^{2}+\Delta^{2}}, ~~~~ \epsilon_{k}=\sqrt{k^{2}+m^{2}}, ~~~~ e=\pm.
\end{equation}

A stable phase must minimize the free energy with respect to the variation of the gap parameter $\partial\Omega_{0} /\partial \Delta = 0$. Then, from (\ref{T0-potential}) we obtain the gap equation
\begin{equation}\label{gap}
1=g\int_{\Lambda}\frac{d^{3}k}{(2\pi)^{3}}\left[\frac{1}{2\epsilon_{k}^{+}}+\frac{1}{2\epsilon_{k}^{-}}\right]
\end{equation}

As usual in the study of the BCS-BEC crossover we will consider a canonical ensemble where the particle number density, $n_{F}=-\partial\Omega_{0}/\partial\mu$, is fixed through the Fermi momentum $P_F$ as $n_{F}=P_{F}^{3}/{3\pi^{2}}$. Then, from (\ref{T0-potential}) we get
\begin{equation}\label{neutrality}
\frac{P_{F}^{3}}{3\pi^{2}}=-\int_{\Lambda}\frac{d^{3}k}{(2\pi)^{3}}\left[\frac{\xi_{k}^{+}}{\epsilon_{k}^{+}}-\frac{\xi_{k}^{-}}{\epsilon_{k}^{-}}\right], \quad \xi_{k}^{\pm}=\epsilon_{k}\mp\mu
\end{equation}

Now, we solve numerically the system of Eqs.~(\ref{gap})~and~(\ref{neutrality}) to find the gap $\Delta$ and chemical potential $\mu$, as a function of the coupling constant $g$. As we will show, varying the strength of $g$ yields the crossover from BCS (for a weak $g$) to BEC (for a strong $g$). We scale the theory parameters so to guarantee a relativistic regime $P_{F}/{\Lambda}=0.3, ~ m/{\Lambda}=0.2$. The results for $\mu$ as functions of $g$, in the interval $0.06>\widetilde{g}>2$, with $\widetilde{g}=g\Lambda^2/4\pi^2$, and for $\Lambda=602.3$ MeV, are shown in Figs. 1.

\begin{figure}
\begin{center}
\includegraphics[width=0.4\linewidth, angle=-90]{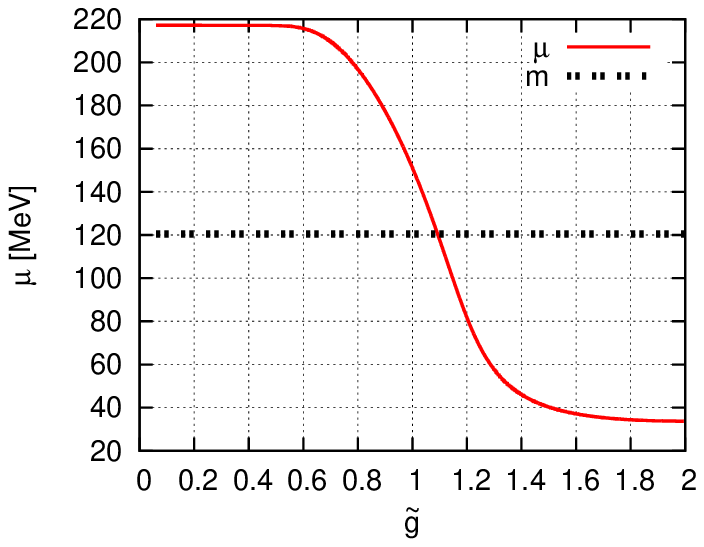}
\caption{Chemical potential, $\mu$, vs $\widetilde{g}=g\Lambda^2/4\pi^2$, and mass $m$.}
\label{fig:mu}
\end{center}
\end{figure}

\begin{figure}
\begin{center}
\includegraphics[width=0.4\linewidth, angle=-90]{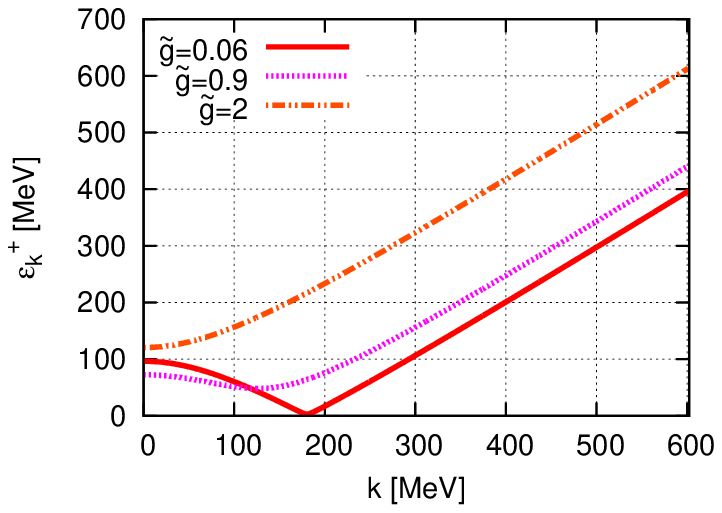}
\caption{$\epsilon_{k}^{+}$ vs $k$ plotted for different $\widetilde{g}=g\Lambda^2/4\pi^2$ values.}
\label{fig:spectrum}
\end{center}
\end{figure}

As known, the condition $\mu < m$ is characteristic of a relativistic Bose gas \cite{BEC-Rel}. From Fig. 1 we see that for this simple model there exists a critical value for the coupling constant $\widetilde{g}_{cr}\sim 1.1$ beyond which the condition $\mu < m$ is satisfied. Thus, we expect to have for $\widetilde{g}>\widetilde{g}_{cr}$ a qualitative change in the properties of the system quasiparticles' modes. Specifically, the quasiparticle spectrum corresponding to coupling constants smaller and larger than $\widetilde{g}_{cr}$ should correspond to fermion-like and boson-like behaviors, respectively. In Fig. 2, we have plotted the quasiparticle spectra, $\epsilon_k^+$, corresponding to different values of the coupling constant. The gap, $\Delta$, and chemical potential, $\mu$, entering in the quasiparticle spectrum (\ref{spectrum}) are obtained as solutions of Eqs. (\ref{gap}) and (\ref{neutrality}) for each value of $\widetilde{g}$. From their graphical representations in Fig. 2, we can see that for the spectra corresponding to $\widetilde{g}=0.06$ and $0.9$ the minimum of their dispersion relations
occurs at $k=\sqrt{\mu^{2}-m^{2}}$, with
excitation energy given by the gap $\Delta$, a behavior
characteristic of quasiparticles in the BCS regime. On the other hand, for $\widetilde{g}=2$, the minimum of the corresponding spectrum occurs at $k=0$, with excitation energy $\sqrt{(\mu-m)^{2}+\Delta^{2}}$, which is typical of Bosonic-like quasiparticle. Therefore, it is corroborated that $\widetilde{g}_{cr}$ is the threshold value for the BCS-BEC crossover in this model.

\section{Equation of State at Strong Coupling}\label{section3}

To investigate how the EoS is affected by the BCS-BEC crossover
 it is needed to find the system energy density and pressure as a function of the coupling-constant strength. Therefore, varying the values of $\widetilde{g}$ from $\widetilde{g}<\widetilde{g}_{cr}$ to $\widetilde{g}>\widetilde{g}_{cr}$ we will be able to describe the EoS corresponding to the BCS and BEC regimes respectively.

The energy density and pressure are obtained from the quantum-statistical average of the energy momentum tensor. For an isotropic system, as the one we are considering, the covariant structure of the $\langle T_{\mu \nu}\rangle$ tensor is given as
\begin{equation}\label{energy-momentum}
\frac{T}{V}~\langle T_{\mu \nu}\rangle=(\Omega_{0}+B)g_{\mu \nu}+(\mu n_{F}+TS)u_\mu u_\nu
\end{equation}
where $V$ is the system volume, $T$ the absolute temperature, $S$ the entropy, and $u_\mu$ the medium 4-velocity with value $u_\mu=(1,\overrightarrow{0})$ in the rest frame. In (\ref{energy-momentum}) we introduced the bag constant $B$ to account for the energy difference between the perturbative vacuum and the true one. In that way, we are modeling what occurs in the case of quark matter, where the asymptotically-free phase of quarks forms a perturbative regime (inside a bag) which is immersed in the nonperturbative vacuum. Then, in the energy density, the energy difference between the perturbative vaccum and the true one should be added. Essentially, that is the bag constant $B$ characterizing a constant energy per unit volume associated to the region where the quarks live. From the point of view of the pressure, $B$ can be interpreted as an inward pressure needed to confine the quarks into the bag. In the numerical calculations we will take $B^{1/4}=145$ MeV, which is a value compatible with that found in the MIT model \cite{MIT}.

Hence, the system energy density and pressure, in the zero-temperature limit, are respectively calculated from $\langle T_{00} \rangle$ and $\langle T_{ii}\rangle, i=1,2,3$, respectively as
\begin{equation}\label{eq:ep}
\varepsilon=\Omega_{0}+\mu n_{F}+B, ~~~~ p=-\Omega_{0}-B
\end{equation}

In Fig. 3 it is plotted $\varepsilon$ and $p$ versus the coupling-constant strength $\widetilde{g}$. Notice that the system energy density is increasing with the coupling strength, while the pressure is decreasing up to get negative values at coupling constants corresponding to the BEC regime. The appearance of a negative pressure for the diquark free gas in the BEC region will indicate that the free-diquark system is unstable.

\begin{figure}
\begin{center}
\includegraphics[width=0.4\linewidth, angle=-90]{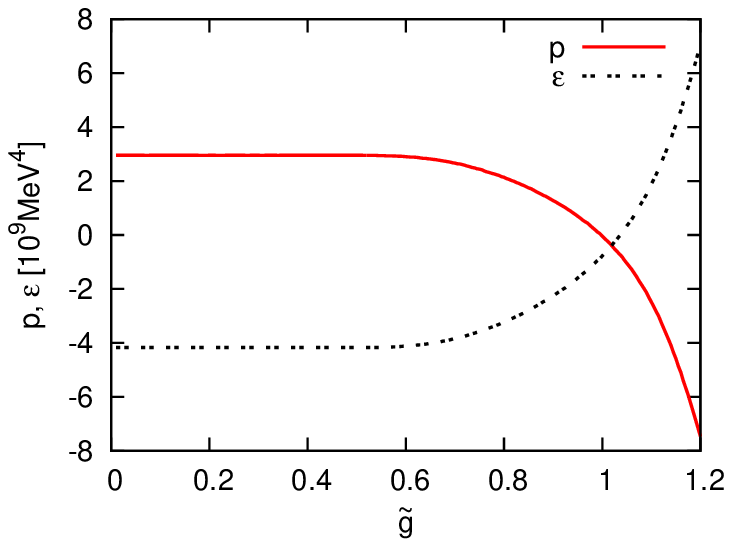}
\caption{Energy density, $\varepsilon$, and pressure, $p$, vs the coupling strength $\widetilde{g}=g\Lambda^2/4\pi^2$ for a free-diquark gas.}
\label{fig:pressureenergy}
\end{center}
\end{figure}

The pressure decay in the BEC region can be explained taking into account the absence of repulsion between the diquarks molecules, once the system is in the BEC region. Nevertheless, as we will show as follows, once we consider the contribution of the diquark-diquark repulsion in the EoS of the strongly interacting system we find that this extra interaction compensates the decreasing tendency due to the Bose-Einstein condensation and consequently rendering a constant pressure throughout the strongly interacting region.

The modeling of self-interacting diquarks in the context of a $\phi^4$ boson theory was initially developed in \cite{Diquarks}. For our fermion system, it can be achieved by introducing a $\lambda \Delta^4$ term in the free energy (\ref{MF-potential})
\begin{equation}\label{MF-potential-2}
\Omega_T=-\frac{1}{\beta}\sum_{n=0}^\infty\int\frac{d^3k}{(2\pi)^4}~Tr \ln~[\beta G^{-1}(i\omega_n,\textbf{k})]+\frac{\Delta^{2}}{g}+\lambda\Delta^4,
\end{equation}

Hence, the system energy density and pressure given in (\ref{eq:ep}) become
\begin{equation}\label{eq:ep-2}
\varepsilon=\Omega_{0}+\lambda\Delta^4+\mu n_{F}+B, ~~~~ p=-\Omega_{0}-\lambda\Delta^4-B
\end{equation}

A possible value for the coupling constant $\lambda$ was estimated as $\lambda=27.8$ in \cite{Diquarks}. It was found taking into account the quark interactions in the context of a modified P-matrix formalism of Jaffe and Low \cite{Jaffe}.

The values for $\Delta$ and $\mu$ obtained for $\lambda=27.8$ from the modified gap equation after including the diquark-diquark repulsive interaction term
\begin{equation}\label{gap-2}
1=g\int_{\Lambda}\frac{d^{3}k}{(2\pi)^{3}}\left[\frac{1}{2\epsilon_{k}^{+}}+\frac{1}{2\epsilon_{k}^{-}}\right]-2\lambda g \Delta^2
\end{equation}
and (\ref{neutrality}), are given in Fig. 4.

\begin{figure}
\begin{center}
\includegraphics[width=0.4\linewidth, angle=-90]{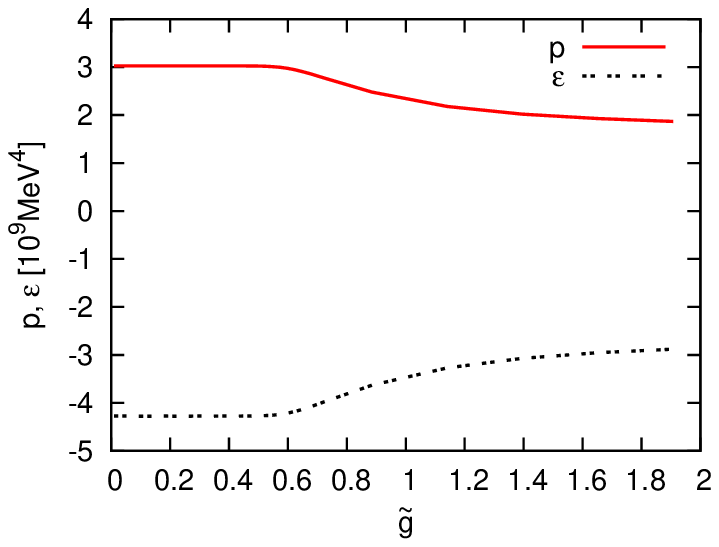}
\caption{Energy density, $\varepsilon$, and pressure, $p$, vs the coupling strength $\widetilde{g}=g\Lambda^2/4\pi^2$ for a self-interacting diquark gas with $\lambda=27.8$.}
\label{fig:pressure-vs-energy}
\end{center}
\end{figure}

The repulsive interaction between diquarks makes a significant contribution to the energy density and pressure (\ref{eq:ep-2}) as can be seen comparing Figs. 3 and 4. From Fig. 4, we see that the matter pressure now remains almost the same in the whole strongly interacting region. This same effect prevents the gas condensation into a zero momentum ground state at zero temperature. In this scenario, the repulsion between diquarks can produce enough outward pressure to elude the star collapse.

On the other hand, as it was shown in \cite{Jason}, once the diquark-diquark repulsion is large enough to oppose the decay of the pressure, then it is obtained that $\mu>m$ for all g-values. This implies the absence of a BEC region. In conclusion, we find that there is no way to put together a BEC dynamics with a positive pressure; meaning that a gravitational-bound compact star cannot be formed by BEC quark molecules.

\section{Magnetic Field Effect on the BCS-BEC Crossover}\label{section4}

To explore the effects of the magnetic field in the crossover, we used
in Ref. \cite{Wang} a model of fermions and scalar bosons interacting via
a Yukawa term, to allow for two
oppositely charged fermions $\Psi^{T}=(\psi_{1},\psi_{2})$ that couple
to an external, uniform and constant magnetic field B. The charged fermions in our model mimic the rotated
charged quarks that pair to form neutral Cooper pairs in the CFL and
2SC phases. The theory is described by the Lagrangian density \begin{equation}
{\cal L}={\cal L}_{f}+{\cal L}_{b}+{\cal L}_{I},\label{Lagrangian}\end{equation}
 with \begin{subequations} \begin{equation}
{\cal L}_{f}=\overline{\Psi}(i\gamma^{\mu}\partial_{\mu}+\mu\gamma^{0}-\widehat{Q}\gamma^{\mu}A_{\mu}-m)\Psi,\end{equation}
 \begin{equation}
{\cal L}_{b}=(\partial_{\mu}+2i\mu\delta_{\mu0})\varphi^{\ast}(\partial^{\mu}-2i\mu\delta^{\mu0})\varphi-m_{b}^{2}\varphi\varphi^{\ast},\end{equation}
 \begin{equation}
{\cal L}_{I}=\varphi\overline{\Psi}_{C}(i\gamma_{5}\widehat{G})\Psi+\varphi^{\ast}\overline{\Psi}(i\gamma_{5}\widehat{G})\Psi_{C}.\end{equation}
 \label{eq2} \end{subequations}Here $m$ and $m_{b}$ denote the
fermion and boson masses respectively. The charge conjugate fermions
are described by $\Psi_{C}=C\overline{\Psi}^{T}$ with $C=i\gamma^{2}\gamma^{0}$
and the electric charge $\widehat{Q}=q\sigma_{3}$ and Yukawa coupling
$\widehat{G}=g\sigma_{2}$ operators are given in terms of the Pauli
matrices $\sigma_{i}$. $A_{\mu}$ is the vector potential associated
with the external magnetic field B, which, without loss of generality,
can be chosen along the $x_{3}$ axis.

The Lagrangian (\ref{Lagrangian})
is invariant under the group of transformations $\mathrm{U(1)_{B}}\otimes\mathrm{U(1)_{em}}$, with subscripts
'B' and 'em' labeling the groups of baryonic and electromagnetic transformations
respectively. In particular the transformation associated with the
$\mathrm{U(1)_{B}}$ symmetry, $\Psi\rightarrow\Psi'=e^{-i\alpha}\Psi$,
$\varphi\rightarrow\varphi'=e^{i2\alpha}\varphi$, implies that the bosons
carry twice the baryon number of the fermions. As a consequence, the
chemical equilibrium with respect to the conversion of two fermions
into one boson and vice versa is ensured by introducing a baryonic chemical
potentials $\mu$ for fermions and $2\mu$ for bosons.

The BCS-BEC crossover can be described as a transition from a regime formed by weakly coupled and neutral Cooper pairs
of two fermions with opposite electric charges to the one formed by molecular difermionic
bound states of electrically neutral boson. In order to describe the BEC of
bosons, we have to separate the zero-mode of the boson field $\varphi$
and replace it by its expectation value $\phi\equiv\langle\varphi\rangle$,
which represents the electric neutral difermion condensate. The mean-field
effective action is then \begin{eqnarray}
I^{B}(\overline{\psi},\psi) & = & \frac{1}{2}\int d^{4}x\, d^{4}y\,\overline{\Psi}_{\pm}(x){\cal S}_{(\pm)}^{-1}(x,y)\Psi_{\pm}(y)
 +(4\mu^{2}-m_{b}^{2})\mid\phi\mid^{2}\nonumber \\
 &  & +\mid(\partial_{t}-2i\mu)\varphi\mid^{2}-\mid\nabla\varphi\mid^{2}-m_{b}^{2}\mid\varphi\mid^{2},\label{b-action}\end{eqnarray}
where the fermion inverse propagators of the Nambu-Gorkov positive
and negative charged fields $\Psi_{+}=(\psi_{2},\psi_{1C})^{T}$ and
$\Psi_{-}=(\psi_{1},\psi_{2C})^{T}$ are given by \begin{equation}
{\cal S}_{(\pm)}^{-1}=\left(\begin{array}{cc}
[G_{(\pm)0}^{+}]^{-1} & i\gamma^{5}\Delta^{*}\\
i\gamma^{5}\Delta & [G_{(\pm)0}^{-}]^{-1}\end{array}\right)\ ,\label{inv-propg}\end{equation}
 with \begin{equation}
[G_{(\pm)0}^{\pm}]^{-1}(x,y)=[i\gamma^{\mu}\Pi_{\mu}^{(\pm)}-m\pm\mu\gamma^{0}]\delta^{4}(x-y)\ ,\label{B-x-inv-prop}\end{equation}
 and $\Pi_{\mu}^{(\pm)}=i\partial_{\mu}\pm qA_{\mu}$ is the covariant derivative. We take the
external vector potential in the Landau gauge $A_{2}=Bx_{1}$, $A_{0}=A_{1}=A_{3}=0$.
The relation between the Bose condensate $\phi$ and the difermion
condensate is given through $\Delta=2g\phi$.

The zero temperature effective potential obtained from (\ref{b-action}) becomes \cite{Wang},

 \begin{equation}
\Omega=-\frac{qB}{2\pi^{2}}\sum_{e=\pm1}\sum_{k=0}^{\infty}g(k)\int_{0}^{\infty}dp_{3}\epsilon_{e}\exp^{-(p_{3}^{2}+qBk)/\Lambda^{2}}
+F_{m_b}(\Delta)+\frac{1}{4\pi^{2}}\sum_{e=\pm1}\int_{0}^{\infty}\omega_{e}p^{2}\, dp,
\label{Eff-Pot}\end{equation}
 where
 \begin{equation}
\epsilon_{e}(k)=\sqrt{(\epsilon_{k}-e\mu)^{2}+\Delta^{2}},\qquad e=\pm1,\quad \epsilon_{k}=\sqrt{p_{3}^{2}+2|q|Bk+m^{2}},\quad k=0,1,2,...
\end{equation}\label{spectrum-f}
and
\begin{equation}
\omega_{e}=\sqrt{p^{2}+m_{b}^{2}}-2e\mu,\qquad e=\pm1, \quad F_{m_b}(\Delta)= \frac{(m_{b}^{2}-4\mu^{2})\Delta^{2}}{4g^{2}}\label{spectrum-b}
\end{equation}
 The index $k$ denotes the Landau levels, $g(k)=[1-(\delta_{k0}/{2})]$ is the spin degeneracy of the Landau levels (LL's) with $k\geq 1$ and $e$ labels quasiparticle/antiquasiparticle
contributions. In order to have only continuous quantities, we introduced in (\ref{Eff-Pot}) a smooth ultraviolet cutoff depending on $\Lambda$ (with
$\Lambda=1$ GeV.).

To investigate the crossover in this case we need to solve the gap equation and the condition
of chemical equilibrium at fixed parameters, and then use them to
obtain the density fractions of fermions and bosons as functions of
the field. Chemical equilibrium requires $n=n_{F}+n_{0}$,
where $n$ plays the role of a fixed total baryon number density, $n=-\partial\Omega/\partial\mu$,
and the fermion number density $n_{F}$ and condensate density $n_{0}$,
are respectively given by

\begin{equation}
n_{F}  =  -\frac{qB}{4\pi^{2}}\sum_{e=\pm1}\sum_{k=0}^{\infty}eg(k)\int_{0}^{\infty}dp_{3}\frac{\epsilon_{k}-e\mu}{\epsilon_{e}}\exp^{-(p_{3}^{2}+qBk)/\Lambda^{2}},\quad
n_{0}  =  \frac{2\mu\Delta^{2}}{g^{2}}.\label{Condensate-density}
\end{equation}

The gap equation is given by $\partial\Omega/\partial\Delta=0$,
which can be obtained from (\ref{Eff-Pot}) as

\begin{equation}\label{Gap-Eq}
\frac{\widetilde{m}_b^2-4\mu^2}{2g^2}=
\frac{qB}{2\pi^2}\sum_{e=\pm 1}\sum_{k=0}^\infty g(k)\int_0^\infty
 \frac{dp_3}{\epsilon_{e}(k)}\exp^{-(p_{3}^{2}+qBk)/\Lambda^{2}}
-\frac{2}{(2\pi)^{3}}\int_{-\infty}^\infty  \frac{d^{3}p}{\sqrt{p^{2}+m^{2}}}.
\end{equation}

As discussed in \cite{Q-Wang}, the crossover parameter in the
present case can be defined by
$x\equiv-\frac{\widetilde{m}_{b}^{2}-4\mu^{2}}{2g^{2}}$, which is
linked to the renormalized boson mass $\widetilde{m}_{b}$ in vacuum
\begin{eqnarray}\label{Boson-mass}
\widetilde{m}_b^2=m_b^2-4g^2\int_{-\infty}^\infty \frac{d^{3}p}{(2\pi)^{3}}
\frac{1}{\sqrt{p^{2}+m^{2}}}.
\end{eqnarray}
The parameter $x$ can then be changed by hand to mimic the
effect of a change in the coupling. Following the derivations of \cite{Q-Wang}, one can
see that at zero magnetic field the parameters of the theory $g$,
$n$, $m$, and $\widetilde{m}_{b}$ can be always chosen to have $x=0$
coinciding with the situation where the density fractions of
fermions $\rho_{F}=n_{F}/n$, and bosons $\rho_{b0}=n_{0}/n$ are all
equal to $1/2$. With such a choice, negative values of $x$ with
large modulus describe a pure BCS state, while large positive values
of $x$ describe a pure BEC phase, and $1/x$ plays the role of the
scattering length. The selection of the model parameters can be done
at any given magnetic field value, to have $x=0$ corresponding to
the unitarity limit, at which the scattering length becomes
infinite. For our purpose, we are more interested in exploring the
situation where we keep fixed values of the parameters, and instead
change the strength of the magnetic field to see if it can have any
effect in the fractions of fermion and boson numbers, and hence
in the BCS-BEC crossing. Henceforth we will use $m=0.2$ GeV, and $g=1$.

What we found in \cite{Wang} is that changing the magnetic field the crossover can be tuned making the system to pass from one region to other. The origin of this effect is in the change with varying fields of the number of quasiparticles with bosonic and fermionic degrees of
freedom. This can be understood in terms of the behavior of
the quasiparticle dispersion relations (\ref{spectrum-f}). Let us introduce the LL-dependent mass square
$M_{k}^{2}\equiv2|q|kB+m^{2}$ in terms of which the quasiparticle dispersion becomes $\epsilon_{+}(k)=\sqrt{(\sqrt{p^{2}_{3}+M_{k}^{2}}-\mu)^{2}+\Delta^{2}}$.
Notice that for all the LLs satisfying the condition
$\mu>M_{k}$, the minimum of the dispersion
$\epsilon_{+}(k)$ occurs at $p_{3}=\sqrt{\mu^{2}-M_{k}^{2}}$, with
excitation energy given by the gap $\Delta$, a behavior
characteristic of the BCS regime. On the other hand, for LLs
with $\mu<M_{k}$, the minimum of $\epsilon_{e}(k)$ occurs at
$p_{3}=0$, with excitation energy $\sqrt{(\mu-M_{k})^{2}+\Delta^{2}}$, typical of the BEC regime.
Therefore, the BCS-BEC crossover in the presence of the magnetic field is controlled by the relative numbers of LLs for which the sign of the effective chemical potential $\mu_{k}=\mu-M_{k}$ is either positive (BCS-type) or negative (BEC-type). Notice, that although we can have $\mu >m$, the presence of the magnetic field can make $\mu < M_k$ for $k\geq 1$.
At fields large enough to put all the fermions in the lowest Landau level (LLL), one has $M_{k}=m$ and the dispersion reduces to $\epsilon_{+}(0)=\sqrt{(\sqrt{p^{2}_{3}+m^{2}}-\mu)^{2}+\Delta^{2}}$, thus the system is in the BCS regime, as long as $\mu>m$.
We call attention that on a close look, the essence of the above description of a field-induced relativistic BCS-BEC crossover is not too different from the essence of the crossover at zero field previously discussed at zero field.

\section{Conclusions}
In this talk I have shown in a simplified model the fact that when the system crossovers from BCS to BEC, the diquark-diquark repulsion is required to make the system stable (i.e. to have a positive pressure); and that, once the diquark-diquark repulsion is included the diquark system loses its BEC nature. On the other hand, we pointed out that a magnetic field could alter the BCS-BEC crossover. Changing the magnetic field strenth the system can corossover from BCS to BEC and viceversa.

I finish by indicating that the presented results should be studied in more realistic models of color superconductivity and that the EoS in the strong-coupling regime should be studied in the presence of a magnetic field. Also the effect of the diquark-diquark repulsion to the Mass-Radio relationship should be investigated.

\bigskip
I want to acknowledge  the CSQCD3 Organizer Committee for their support and hospitality in Guarujá, Sao Paulo, Brazil. This work has been supported in part by the Office of Nuclear Theory of
the Department of Energy under contract de-sc0002179.


\begin{thebibliography}{99}

\bibitem{Fukushima}K.  Fukushima and T. Hatsuda, Rept. Prog. Phys. 74, 014001 (2011).

\bibitem{Politzer}
  H. D. Politzer, Phys. Rev. Lett. 30, 1346 (1973); D. J. Gross, and F. Wilczek, Phys. Rev. Lett. 30, 1343 (1973).

\bibitem{CFL} M. Alford, K. Rajagopal and F. Wilczek, Phys. Lett. B 422, 247 (1998); R. Rapp, T. Schafer,
E. V. Shuryak, and M. Velkovsky, Phys. Rev. Lett. 81, 53 (1998).

\bibitem{gCFL}M. Alford, C. Kouvaris, and K. Rajagopal, Phys. Rev. Lett. B 92, 222001 (2004); Phys. Rev. D 71, 054009 (2005).

\bibitem{Strong-Coupling} S. B. R\"{u}ster, \textit{et. al}, Phys. Rev. D 72, 034004 (2005); M. Kitazawa, D. H. Rischke and I. A. Shovkovy, Phys. Lett. B 637, 367 (2006); H. Abuki and T. Kunihiro, Nucl. Phys. A 768, 118 (2006).

\bibitem{Coher-length} M. Matsuzaki, Phys. Rev. D 62, 017501 (2000); H. Abuki, T. Hatsuda and K. Itakura, Phys. Rev. D 65, 074014 (2002); K. Itakura, Nucl. Phys. A 715, 859 (2003).

\bibitem{BCS-BEC}Q. Chen, et al., Phys. Rep. 412, 195 (2005).




\bibitem{Jason} E. J. Ferrer and J. P. Keith, Phys. Rev. C 86, 035205 (2012).


\bibitem{Wang} J-c. Wang, V. de la Incera, E. J. Ferrer, Q. Wang, Phys. Rev. D 84, 065014 (2011).

\bibitem{Pulsars} J.~H.~Taylor \textit{et al}., Astrophys. J. S 88,
529 (1993); A.~G.~Lyne, F.~Graham-Smith, \textit{Pulsar Astronomy}
(Cambridge Univ. Press., Cambridge, 2005).

\bibitem{Magnetars} B.~Paczynski, Acta Astron. 42, 145
(1992); C.~Thompson and R.~C.~Duncan, ApJ. 392, L9 (1992);
473, 322 (1996); A.~Melatos, Astrophys. J. Lett. 519,
L77 (1999).

\bibitem{virial}L.~Dong and S.~L.~Shapiro ApJ. 383,
745 (1991).

\bibitem{EoS-B2010} E.~J.~ Ferrer \textit{et al}., Phys. Rev. C
82, 065802 (2010).

\bibitem{Kharzaeev}D.E. Kharzaeev, L.D. McLerran and H.J. Warringa, Nucl. Phys. A 803, 227 (2008).

 \bibitem{Shokov}V.V. Shokov, A.Yu. Illarionov and V.D. Toneev, Int.  J. Mod. Phys. A24, 5925 (2009); V. Voronyuk, et al., Phys. Rev. C 83, 054911 (2011).

\bibitem{Abelev}B.I. Abelev et al. [STAR Collaboration], Phys. Rev. Lett. 103, 251601 (2009); B.I. Abelev et al. [STAR Collaboration],  arXiv:0909.1717 [nucl-ex].


\bibitem{Larry}L.D. McLerran and R. Venugopalan, Phys. Rev. D 49, 2233 (1994); 3352; D 50, 2225 (1994).

\bibitem{R-BCS-BEC-1} Y.~Nishida and H.~Abuki, Phys. Rev. D 72, 096004 (2005); L. He and P. Zhuang, Phys. Rev. D 75, 096003 (2007); 76. 056003 (2007).

\bibitem{BEC-Rel}H. E. Haber and H. A. Weldon, Phys. Rev. Lett. 46, 1497 (1981).




\bibitem{MIT}A. Chodos et al., Phys. Rev. D 9, 3471 (1974); 10, 2599 (1974); T. D. Grand, Phys. Rev. D 12, 2060
(1975).

\bibitem{Diquarks} J. F. Donoghue and K. S. Sateesh, Phys. Rev. D 38, 360 (1988).


\bibitem{Jaffe} R. L. Jaffe and F. E. Low, Phys. Rev. D 19, 2105 (1979).

\bibitem{Q-Wang} J.~Deng, A.~Schmitt and Q.~Wang, Phys. Rev. D
76, 034013 (2007).


\end{thebibliography}
\end{document}